\documentclass[10pt]{article}

\usepackage[a4paper,margin=1in]{geometry}
\usepackage{titling}
\usepackage{authblk}
\usepackage{amsmath,amssymb}
\usepackage{graphicx}
\usepackage[numbers,sort&compress]{natbib}
\usepackage{hyperref}

\usepackage{titling}
\title{Acoustic radiation force on a liquid particle in a standing surface acoustic wave field}

\author[1]{Shuo Huang}
\author[2]{Hemin Pan}
\author[2]{Daniel Ahmed}
\author[1]{Thierry Baasch}

\affil[1]{Department of Biomedical Engineering, Lund University, Lund, Sweden}
\affil[2]{ARTORG Center for Biomedical Engineering Research, Faculty of Medicine, University of Bern, CH-3008 Bern, Switzerland}

\date{} 

\begin{document}
\maketitle

\begin{abstract}
We develop a theory for the acoustic radiation force on a liquid particle in a 2D standing-wave field beyond the Rayleigh limit. The theory is valid for any frequency, includes the traveling-wave components due to the Rayleigh angle, and is thus applicable to a large class of surface acoustic wave applications. The analytical results are validated with respect to finite-element models. Using our analytical solution, we determine the parameter space for which Rayleigh-limit methods, such as the Gor'kov framework, remain applicable. This range is shown to depend on the particle properties, the Rayleigh angle, and even the particle position in the acoustic field. We propose a general form for the acoustophoretic contrast factor applicable to any wavelength of 1D standing-wave field, broadening the applicability of the classical Gor'kov framework. We show that the Rayleigh-angle effect can substantially weaken the acoustic radiation force, an effect that has been largely overlooked. We also confirm a frequency-dependent topological transition of the acoustic landscape that induces a switching of the field attractors and particle equilibrium points. These results advance the quantitative theory of acoustic forces, unveil previously unresolved dynamical features of acoustofluidic fields, and provide a theoretical foundation for SAW-based cell trapping, separation, and enrichment in acoustofluidics.
\end{abstract}


\section{Introduction}
\label{sec:headings}

As a major branch of microfluidics, acoustofluidics uses externally applied acoustic fields for the contactless manipulation of micro- and nanoparticles~\citep{rufo2022acoustofluidics,lenshof2012acoustofluidics}, which holds broad promise for biomedical and chemical applications~\citep{li2018applications,ohlsson2018acoustic,liu2019lipid,magnusson2024acoustic}. Although particles in acoustofluidic systems are subjected to complex acoustic and hydrodynamic effects~\citep{pavlivc2025acoustic,durlofsky1987dynamic}, the main driving effect is the acoustic radiation force~\citep{muller2012numerical,hahn2015numerical,augustsson2016iso}. The acoustic radiation force is a second-order time-averaged effect that originates from non-linearities in the Navier-Stokes equations~\citep{king1934acoustic,yosioka1955acoustic,gor1962forces}.

The most commonly applied framework for calculating the acoustic radiation force is the approach introduced by Gor'kov~\citep{gor1962forces}. Here, the force is obtained directly from the negative gradient of a force potential, commonly referred to as Gor'kov potential. Its key limitation is that it retains only the two lowest-order terms in the multipole expansion, namely the monopole and dipole contributions. As a consequence, the Gor'kov framework is only applicable for wavelengths much larger than the particle diameter, i.e. within the Rayleigh limit. 

In surface-acoustic-wave (SAW) systems, which are widely used in acoustofluidics~\citep{ding2012chip,yeo2014surface,guo2015controlling}, the operating frequency typically lies in the range of 10-100 MHz. Such high frequencies result in wavelengths between 150 - 15 $\mu$m, which are close to the typical sizes of biological cells. Under such conditions, the Gor'kov potential may not yield accurate results. In addition, during the propagation of SAWs from the bottom substrate into the fluid, refraction occurs as a consequence of the phase-matching condition across the fluid--solid interface. As a result, the surface waves that originally propagate along the substrate surface of the SAW device enter the fluid at an inclination angle~\citep{stringer2023methodologies}. This effect is referred to as the Rayleigh-angle effect, which can significantly influence the acoustic radiation force~\citep{liang2018radiation,riaud2020mechanical, peng2020standing}. On the theoretical side, several models have been developed for calculating the acoustic radiation force beyond the Rayleigh limit by retaining higher-order scattering contributions. Silva et al.~\citep{silva2019particle} presented force fields for solid particles in two-dimensional (2D) standing-wave fields, but their theory neither applies to liquid particles such as cells nor does it include the Rayleigh-angle effect. The theory of Liang et al.~\citep{liang2018radiation} is likewise limited to solid particles and restricted to 1D standing waves. A similar 1D restriction also appears in Marston's studies of acoustic radiation forces based on partial-wave-series expansion~\citep{marston2024contrast}. Although the theory developed by Doinikov ~\citep{doinikov2001acoustic} is mathematically complete, the work does not include the specific results for particles immersed in 2D wave-fields. Nevertheless, it is undeniable that high-frequency operation and the Rayleigh-angle effect lie at the core of many SAW-based acoustofluidic techniques for particle and cell manipulation, such as one-cell-per-well (OCPW) trapping~\citep{collins2015two}, tilted-SAW-based cell separation~\citep{ding2014cell}, three-dimensional cell manipulation~\citep{guo2016three}, and the control of cell--cell interactions~\citep{guo2015controlling}.

Motivated by these limitations, this study focuses on two main objectives. First, we develop a theoretical framework for the acoustic radiation force on liquid particles, such as cells, incorporating acoustic scattering up to any order and the Rayleigh-angle effect. Second, based on this framework, we carry out a systematic analysis of the dynamical characteristics of 2D standing-wave fields widely used in SAW devices. We map the conditions under which higher-order modes and the influence of the Rayleigh-angle on the acoustic radiation force are significant. Notably, the present theory yields a general expression for the acoustophoretic contrast factor in a 1D standing-wave field. We also identify a topological transition of the acoustic potential landscape, revealing how the attractors and particle equilibrium points in the acoustic field switch as a function of the frequency and particle physical properties. The present work advances the fundamental theory of acoustophoresis and holds considerable practical value for applications involving cell trapping, separation, and manipulation.

\section{Theory}

In this section, we develop a theoretical framework for the acoustic radiation force exerted on a spherical liquid particle suspended in high-frequency 2D standing-wave fields beyond the Rayleigh limit. The solution includes the Rayleigh-angle effect arising from the refraction and transmission of surface acoustic wave at fluid–solid interfaces. This framework is hereafter referred to as the RA framework.

\subsection{Problem statement}

Here we aim to calculate the acoustic radiation force on a liquid particle of unperturbed density $\rho_p$ and compressibility $\kappa_p$ immersed in a fluid of unperturbed density $\rho_0$ and speed of sound $c_0$. As the acoustic radiation force depends on the incident and scattered pressure sound fields, we will solve the scattering problem. The dynamics of pressure fluctuations are then well described by the wave equation. Neglecting transients and assuming time-harmonic perturbations of angular frequency $\omega$ reduces the wave equation to the Helmholtz equation, yielding

\begin{align}\label{eq:Helmholtz}
\omega^2 p +& c_0^2 \nabla^2 p =0 \text{ in the fluid domain},  \\
\omega^2 p +& c_p^2 \nabla^2 p =0  \text{ in the particle domain},
\end{align}
where $c_p^2=1/(\rho_p \kappa_p)$. The instantaneous pressure is given by $\text{ Re} \left[  p {\rm e}^{- {\rm i} \omega t} \right] $.
The relation between acoustic pressure and fluid velocity is then given by 
\begin{align}
&\rho_0 \frac{{\partial v}}{{\partial t}} = - \nabla p\text{ in the fluid domain},  \\
&\rho_p \frac{{\partial v}}{{\partial t}} = - \nabla p \text{ in the particle domain}.
\end{align}
Noting that the present study focuses on the liquid particle, the boundary conditions at the particle surface require the continuity of pressure and normal velocity across the interface:
\begin{equation}
\left\{
\begin{array}{l}
p_{in}+p_{sc}=p_{lq}, \\[4pt]
v_{in}+v_{sc}=v_{lq},
\end{array}
\right.
\label{eq:continuity conditions mother}
\end{equation}
in which $p_{in}$, $p_{sc}$, and $p_{lq}$ denote the pressure associated with the incident wave, the scattered wave, and the acoustic wave inside the liquid particle, respectively, and $v_{in}$, $v_{sc}$, and $v_{lq}$ denote the corresponding normal velocities at the particle interface. In this set of equations $p_{in}$ defines the problem and will be introduced in Section \ref{sec:beamshape}.

\subsection{Acoustic radiation force}

We consider a right-handed Cartesian coordinate system ($x$, $y$, $z$), in which four plane traveling waves of identical frequency $f$ are present. Two of these waves propagate in opposite directions along the $x$-axis, while the other two propagate oppositely along the $z$-axis. The superposition of these four waves gives rise to a 2D standing-wave field in the $x$-$z$ plane. The pressure amplitude of the standing wave in the $z$-direction is denoted by $p_{0z}$. The pressure amplitude of the standing wave in the $x$-direction is correspondingly written as $\epsilon p_{0z}$, where $\epsilon$ is a dimensionless scaling factor accounting for the amplitude imbalance between the two directions. Under such an acoustic field, the primary acoustic radiation force acting on a particle can be expressed as~\citep{silva2019particle}:
\begin{equation}
  \mathbf{F} = 2\pi a^2 E_{ac} \mathbf{Q}_{\mathrm{rad}}.
\label{eq:mother equation}
\end{equation}
Here, $a$ denotes the particle radius, and $E_{ac} = \frac{{p_{0z}^2}}{{4{\rho _0}c_0^2}}$ represents the characteristic acoustic energy density, where $\rho_0$ and $c_0$ are the density and sound speed of the background fluid, respectively. The
radiation force acting on a particle can be expressed as:
\begin{equation}
  \mathbf{Q_{rad}} = ({Q_x}\mathbf{e_x} + {Q_z}\mathbf{e_z}),
\label{eq:Q mother}
\end{equation}
where $\mathbf{Q_{rad}}$ denotes the radiation-force efficiency vector, whose Cartesian components can be expressed as
\begin{equation}
\begin{array}{l}
{Q_x} =  - \frac{1}{{2\pi {{(ka)}^2}}}{\mathop{\rm Im}\nolimits} \sum\nolimits_{n,m} {\sqrt {\frac{{(n + m + 1)(n + m + 2)}}{{(2n + 1)(2n + 3)}}} } ({S_n}{a_{nm}}a_{n + 1,m + 1}^* + S_n^*{a^*_{ n,-m}}a_{n + 1, - m - 1}),\\
{Q_z} = \frac{1}{{\pi {{(ka)}^2}}}{\mathop{\rm Im}\nolimits} \sum\nolimits_{n,m} {\sqrt {\frac{{(n - m + 1)(n + m + 1)}}{{(2n + 1)(2n + 3)}}} } {S_n}{a_{nm}}a_{n + 1,m}^*.
\end{array}
\label{eq:Qx and Qy}
\end{equation}
In this expression, $\sum_{n,m}=\sum_{n=0}^{\infty}\sum_{m=-n}^{n}$, and $a_{nm}$ are the beam-shape coefficients (BSCs), which depend on the characteristics of the incident acoustic field, while
\begin{equation}
{S_n} = {s_n} + s_{n+1}^* + 2{s_n}s_{n+1}^*,
\label{eq:S_n}
\end{equation}
and $s_n$ are the scattering coefficients, determined by the properties of the scattered acoustic field. It follows from Eq.~\eqref{eq:Qx and Qy} that once the explicit forms of $a_{nm}$ and $s_n$ are known, the acoustic radiation force acting on a particle can be fully determined.

\subsection{Beam-shape coefficients for the incident field including the Rayleigh angle}
\label{sec:beamshape}

Considering the spherical harmonic expansion, the pressure of a traveling wave with an amplitude of $p_{0z}/2$ can be written as 
\begin{equation}
{p_{tr}} = \frac{p_{0z}}{2}{e^{i\vec k \cdot {{\vec r}_j}}}{e^{i\vec k \cdot {{\vec r}_i}}} = 2\pi {p_{0z}}{e^{i\vec k \cdot {{\vec r}_j}}}\sum\limits_{n,m} {{i^n}Y_n^m(\alpha ,\beta )} {j_n}({k_0}r)Y_n^m(\theta ,\varphi ).
\label{eq:p_in spherical}
\end{equation}
In this expression, ${\vec r}_j$ is the position vector of particle $j$ in the Cartesian coordinate system, ${\vec r}_i$ denotes the position vector of an arbitrary point $i$ in the acoustic field relative to the particle center, $j_n$ denotes the spherical Bessel function of the first kind, and $Y_n$ denotes the spherical harmonic function. $\alpha$ and $\beta$ denote the polar and azimuthal angles of the wave vector of the incident traveling wave, respectively, while $\theta$ and $\varphi$ denote the polar and azimuthal angles of the position vector of point $i$, respectively.

For surface acoustic waves, the Rayleigh angle $\vartheta$ must also be taken into account. This angle depends on the substrate sound speed $c_S$ of the SAW device and the sound speed $c_0$ of the fluid, and satisfies $\vartheta  = \arcsin ({c_0}/{c_S})$. In SAW devices using lithium niobate as the substrate material, the Rayleigh angle is typically about $22^\circ$~\citep{stringer2023methodologies}. When the Rayleigh angle is taken into account, the wave vectors of the four traveling waves can be written as
\begin{equation}
\begin{array}{l}
\left\{ \begin{array}{l}
{{\vec k}_1} = k_0 (\sin \vartheta ,\cos \vartheta ,0),\\
{{\vec k}_2} = k_0( - \sin \vartheta ,\cos \vartheta ,0),\\
{{\vec k}_3} = k_0(0,\cos \vartheta ,\sin \vartheta ),\\
{{\vec k}_4} = k_0(0,\cos \vartheta , - \sin \vartheta ),
\end{array} \right.
\end{array}
\label{eq:k vector}
\end{equation}
where $k_0= \omega / c_0$ denotes the wavenumber in the fluid. 
Starting from the basic expression for a traveling wave given in Eq.~\eqref{eq:p_in spherical} and combining it with the four wave orientations specified in Eq.~\eqref{eq:k vector}, we obtain, by superposition, the governing expression for the pressure field considered in this study:
\begin{equation}
p(x,y,z) = {p_{0z}}{e^{ik_0y\cos \vartheta }}[\varepsilon \cos ({k_0}x\sin \vartheta ) + \cos ({k_0}z\sin \vartheta )].
\label{eq:2D pressure}
\end{equation}
Eq.~\eqref{eq:2D pressure} not only produces a 2D standing-wave field in the $xz$ plane, but also generates a traveling-wave component radiating along the $y$-direction. This shows that the presence of the Rayleigh angle significantly modifies the pressure field and, consequently, the dynamical response of particles subjected to SAWs.

The solution of Eq.~\eqref{eq:Helmholtz} also gives the pressure field $p_{tr}$ of a single traveling wave propagating in the fluid domain, which is equivalent to Eq.~\eqref{eq:p_in spherical}, as expressed in Eq.~\eqref{eq:p_in}:
\begin{equation}
{p_{tr}} = \frac{p_{0z}}{2}\sum\limits_{n,m} {{a_{nm}^{tr}}(x,z){j_n}({k_0}r)} Y_n^m(\theta ,\varphi ),
\label{eq:p_in}
\end{equation}
in which $a_{nm}^{tr}$ is the beam-shape coefficient (BSC) corresponding to a single traveling wave. By comparing Eqs.~\eqref{eq:p_in} and~\eqref{eq:p_in spherical}, and further expanding the spherical harmonic function $Y_n^m(\alpha,\beta)=\sqrt{\frac{(2n+1)(n-m)!}{4\pi(n+m)!}}\,P_n^m(\cos\alpha)e^{-im\beta}$, the BSC can be obtained as
\begin{equation}
a_{nm}^{tr} = {i^n}\sqrt {4\pi (2n + 1)\frac{{(n - m)!}}{{(n + m)!}}} {e^{i\vec k \cdot {{\vec r}_j}}}P_n^m(\cos \alpha ){e^{ - im\beta }},
\label{eq:a_nm general}
\end{equation}
where $P_n^m(\cos \alpha )$ denote the associated Legendre polynomials. 

Based on $a_{nm}^{tr}$ for a traveling wave, our next objective is to determine the BSC corresponding to the 2D standing-wave field including the Rayleigh-angle effect as given by Eq.~\eqref{eq:2D pressure}, which we denote by $a_{nm}$. Starting from Eq.~\eqref{eq:p_in} and superposing the four traveling described by Eq.~\eqref{eq:k vector}, the acoustic field in Eq.~\eqref{eq:2D pressure} can be expressed in another form as
\begin{equation}
p(x,y,z)=\sum\limits_{q = 1}^4({p_{tr}})_q = p_{0z}\sum\limits_{n,m} [\sum\limits_{q = 1}^4 ({\frac{{a_{nm}^{tr}}}{2})_q]{j_n}({k_0}r)} Y_n^m(\theta ,\varphi ).
\label{eq:p_tr_sum}
\end{equation}
It should be noted that, in the above expression, the order of summation can first be interchanged. Moreover, both ${j_n}({k_0}r)$ and $Y_n^m(\theta,\varphi)$ are independent of the propagation direction of the traveling waves. According to Eq.~\eqref{eq:p_tr_sum}, the BSC of $p(x,y,z)$ can be written as
\begin{equation}
\begin{split}
a_{nm} = \sum\limits_{q = 1}^4 {({\frac{a_{nm}^{tr}}{2})_q}} ={}& i^n \sqrt{4\pi(2n+1)\frac{(n-m)!}{(n+m)!}}\, e^{ik_0 y \cos\vartheta}
\Biggl[
\\ &\qquad
\varepsilon P_n^m(0) (-i)^m\,
\frac{e^{ik_0x \sin\vartheta+im\vartheta}+e^{-ik_0x \sin\vartheta-im\vartheta}}{2}
\\ &\qquad
+ P_n^m(\sin\vartheta)\, (-i)^m\,
\frac{e^{ik_0z \sin\vartheta}+(-1)^{n+m} e^{-ik_0z \sin\vartheta}}{2}
\Biggr].
\end{split}
\label{eq:anm}
\end{equation}
The above expression can be further rearranged into a more compact form
\begin{equation}
\begin{aligned}
a_{nm}
&= \sqrt{4\pi(2n+1)\frac{(n-m)!}{(n+m)!}}\,e^{ik_0 y \cos\vartheta}
\Big[
\varepsilon P_n^m(0) i^{n-m} \cos(k_0 x\sin\vartheta + m\vartheta)
\\
&\qquad
+ P_n^m(\sin\vartheta)\cos\!\left(k_0 z\sin\vartheta + \frac{\pi}{2}n - \frac{\pi}{2}m\right)
\Big].
\end{aligned}
\label{eq:a_nm standing}
\end{equation}

\subsection{Acoustic scattering of the particle}
For the scattering coefficients $s_n$, we first express the scattered acoustic field in the surrounding fluid $p_{sc}$, and the acoustic field inside the liquid particle $p_{lq}$, as~\citep{silva2019particle}:
\begin{equation}
\left\{ \begin{array}{l}
{p_{sc}} = \frac{p_{0z}}{2}\sum\limits_{n,m} {{s_n}{a_{nm}}(x,z)h_n^{(1)}({k_0}r)Y_n^m(\theta ,\varphi )}, \\
{p_{lq}} = \frac{p_{0z}}{2}\sum\limits_{n,m} {{c_{nm}}(x,z){j_n}({k_p}r)Y_n^m(\theta ,\varphi )}.
\end{array} \right.
\label{eq:psc and plq}
\end{equation}
Here, $k_p$ is the wavenumber inside the liquid particle. 

Combining the boundary conditions with Eqs.~\eqref{eq:p_in} and~\eqref{eq:psc and plq} yields a linear system of equations associated with the two continuity conditions,

\begin{equation}
\left\{
\begin{array}{l}
s_n h_n^{(1)}(k_0 a)-\dfrac{c_{nm}}{a_{nm}}\, j_n(k_p a)=-j_n(k_0 a), \\[4pt]
s_n {h_n^{(1)}}'(k_0 a)-\dfrac{c_{nm}}{a_{nm}}\dfrac{\rho_0 k_p}{\rho_p k_0}\, j_n'(k_p a)=-j_n'(k_0 a).
\end{array}
\right.
\label{eq:continuity conditions}
\end{equation}
By applying Cramer's rule, the scattering coefficients $s_n$ can be obtained explicitly as
\begin{equation}
{s_n} =  - \frac{{\left| {\begin{array}{*{20}{c}}
{{j_n}({k_0}a)}&{{j_n}({k_p}a)}\\
{{{j'}_n}({k_0}a)}&{\frac{{{\rho _0}{k_p}}}{{{\rho_p}{k_0}}}{{j'}_n}({k_p}a)}
\end{array}} \right|}}{{\left| {\begin{array}{*{20}{c}}
{h_n^{(1)}({k_0}a)}&{{j_n}({k_p}a)}\\
{h{{_n^{(1)}}^\prime }({k_0}a)}&{\frac{{{\rho _0}{k_p}}}{{{\rho_p}{k_0}}}{{j'}_n}({k_p}a)}
\end{array}} \right|}}.
\label{eq:sn}
\end{equation}
Finally, substituting Eqs.~\eqref{eq:a_nm standing}, and~\eqref{eq:sn} into the definition of the radiation-force efficiency vector Eq.~\eqref{eq:Qx and Qy} yields the radiation-force efficiency $\mathbf{Q}$ for a standing-wave field that is nominally two-dimensional but becomes pseudo-three-dimensional in the presence of the Rayleigh angle $\vartheta$. The acoustic radiation force acting on the particle can then be directly computed from Eq.~\eqref{eq:mother equation}.

\subsection{Force components}\label{sec:Q expression}

The expressions for the radiation-force efficiency follow from Eqs.~\eqref{eq:a_nm standing}, and~\eqref{eq:sn}. To focus on the principal results, only the compact expressions are retained as shown in Eq.~\eqref{eq:Q RA}. The definitions of the auxiliary coefficients and their detailed derivations are given in Appendix A.
\begin{equation}
\begin{array}{l}
{Q_x} = \epsilon{Q_\alpha }\cos(k_0z\sin\vartheta)\sin(k_0x\sin\vartheta) + \epsilon^2{Q_\beta }\sin(2k_0x\sin\vartheta),\\
{Q_z} = \epsilon{Q_\gamma }\sin(k_0z\sin\vartheta)\cos(k_0x\sin\vartheta) + {Q_\varphi }\sin(2k_0z\sin\vartheta).
\end{array}
\label{eq:Q RA}
\end{equation}
The radiation-force expressions given by Eq.~\eqref{eq:Q RA} and Appendix A constitute a general formulation for the acoustic radiation force in a 2D standing wave field incorporating the Rayleigh angle. As can be seen from the explicit forms of Eqs.~\eqref{eq:Q RA}, the acoustic radiation force is governed primarily by the parameter $k_0a$ and the Rayleigh angle $\vartheta$. These quantities characterize the frequency scale of the acoustic field, the size scale of the particle, and the acoustic properties of the SAW device.

\section{Results}
In this section, we first validate the theory derived in this study (RA framework) using finite-element results as a reference. In the validation, we set $\epsilon$=1, corresponding to an acoustic field formed by two orthogonal identical standing waves. We then investigate the acoustic radiation force in 1D and 2D cases, map the stability of equilibrium positions, and present an analytical formula for the acoustic contrast factor. 

\subsection{Validation against FEM}
We performed finite-element simulations in COMSOL Multiphysics to evaluate the acoustic radiation force acting on liquid particles, and compared the numerical results directly with the predictions of the RA framework. In these simulations, the acoustic field was computed using the \textit{Pressure Acoustics, Frequency Domain} module in COMSOL. The background fluid was water with sound speed $c_0$ = 1500 m/s and density $\rho_0 = 1000~\mathrm{kg/m^{3}}$. The liquid particle was chosen as a white blood cell (WBC), and lithium niobate was used as the substrate material for the SAW device. Table~\ref{tab:model_parameters} summarizes the parameter ranges considered, including the particle radius $a$~\citep{wu2019acoustofluidic}, particle sound speed $c_p$, particle density $\rho_p$~\citep{cushing2017ultrasound}, and substrate sound speed $c_S$~\citep{sachs2023behavior}.

\begin{table}
  \begin{center}
  \begin{tabular}{@{}ll@{}}
    Particle         & $a$ = 6 $\mu m$ , $c_p = 1554\ \mathrm{m/s}$, $\rho_p = 1054\ \mathrm{kg/m^{3}}$ \\
    Background fluid & $c_0 = 1500\ \mathrm{m/s}$, $\rho_0 = 1000\ \mathrm{kg/m^{3}}$ \\
    Lithium niobate  & $c_S = 3860\ \mathrm{m/s}$
  \end{tabular}
  \caption{Model parameters used in the simulation of COMSOL.}
  \label{tab:model_parameters}
  \end{center}
\end{table}

\begin{figure}[!h]
  \centerline{\includegraphics[width=0.45\textwidth]{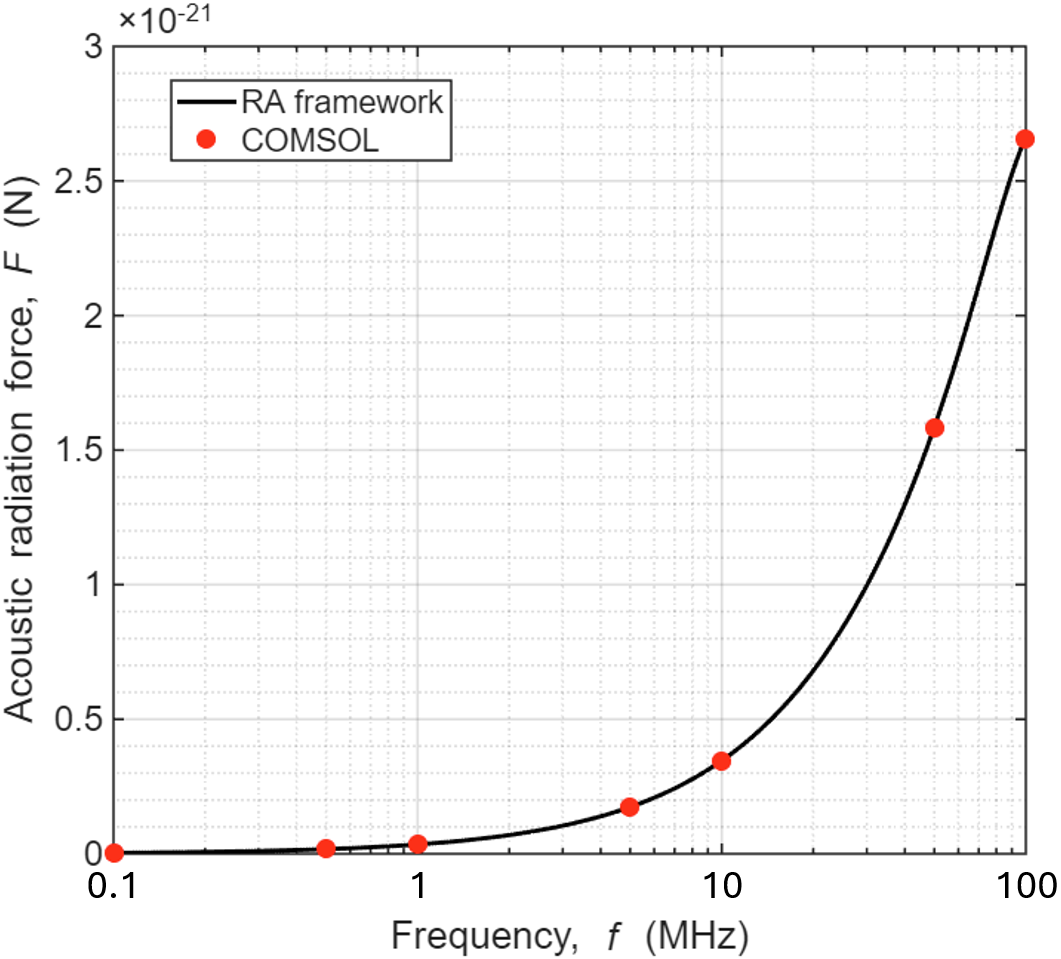}}
  \caption{Comparison between the theoretical predictions and the simulation results for the acoustic radiation force as the frequency of the acoustic field is varied from low to high with the WBC position fixed. The theoretical force curve agrees well with the seven simulation data points, and the maximum relative error is only 0.35\%}
\label{fig:FEM validation} 
\end{figure}

We fixed the particle position at the arbitrarily chosen location $(0.617\lambda_x,\,0.363\lambda_z)$, where $\lambda_x$ and $\lambda_z$ denote the wavelengths in the $x$- and $z$-directions, respectively. Seven frequencies, $f$ = [0.1, 0.5, 1, 5, 10, 50, 100] MHz, were selected from low to high for the simulations, and the corresponding results were compared with the predictions of the RA framework, as shown in Fig.~\ref{fig:FEM validation}.

It can be seen that the predictions of the RA framework agree well with the FEM results obtained in COMSOL over the entire frequency range considered. Among the seven frequencies examined, the largest relative error is 0.35\% at 100 MHz, which is still very small. This demonstrates the validity of the RA framework.

To assess the applicability of the model over a broad parameter range, we fixed the background fluid as water and varied the acoustic frequency $f$, particle radius $a$, particle sound speed $c_p$, particle density $\rho_p$, substrate sound speed $c_S$ of the SAW device, and the particle phase over wide ranges. The detailed parameter settings are listed in Table~\ref{tab:FEM_parameters}. The values given in the table were combined with one another, and, for the particle phase in the acoustic field, four points were randomly sampled within one wavelength in both the $x$- and $z$-directions, yielding a total of 972 cases for comparison.

\begin{table}
  \begin{center}
  \begin{tabular}{ll}
    Acoustic frequency $f$ (MHz)      & 5, 50, 100 \\
    Particle radius $a$ ($\mu$m)      & 0.5, 3, 10 \\
    Particle sound speed $c_p$ (m/s)  & 1700, 2000, 2300 \\
    Particle density $\rho_p$ (kg/m$^3$) & 1050, 1150, 1250 \\
    Substrate sound speed $c_s$ (m/s) & 2500, 3250, 4000
  \end{tabular}
  \caption{Physical parameters used in FEM simulations.}
  \label{tab:FEM_parameters}
  \end{center}
\end{table}

Across all 972 validation cases, the proposed theory shows excellent agreement with the FEM results. The mean relative error is $-1.50\times10^{-3}$, with a standard deviation of $1.67\times10^{-2}$ and an RMSE of $1.68\times10^{-2}$, indicating that the overall discrepancy remains small. Moreover, $90.64\%$, $93.93\%$, and $96.91\%$ of the samples have absolute relative errors below $1\%$, $3\%$, and $5\%$, respectively, confirming the high predictive accuracy of the present model. The remaining errors are likely attributable to the mesh discretization used in the finite-element calculations. Because a finer mesh would substantially increase the computational cost, we did not pursue a higher level of numerical accuracy in the present validations. Overall, these results demonstrate that the present theory provides highly accurate predictions over the vast majority of the parameter space, with only a few outlying cases showing relatively larger errors.

\subsection{1D standing waves: contrast factor with high-order harmonics}

\begin{figure}[!h]
  \centerline{\includegraphics[width=0.9\textwidth]{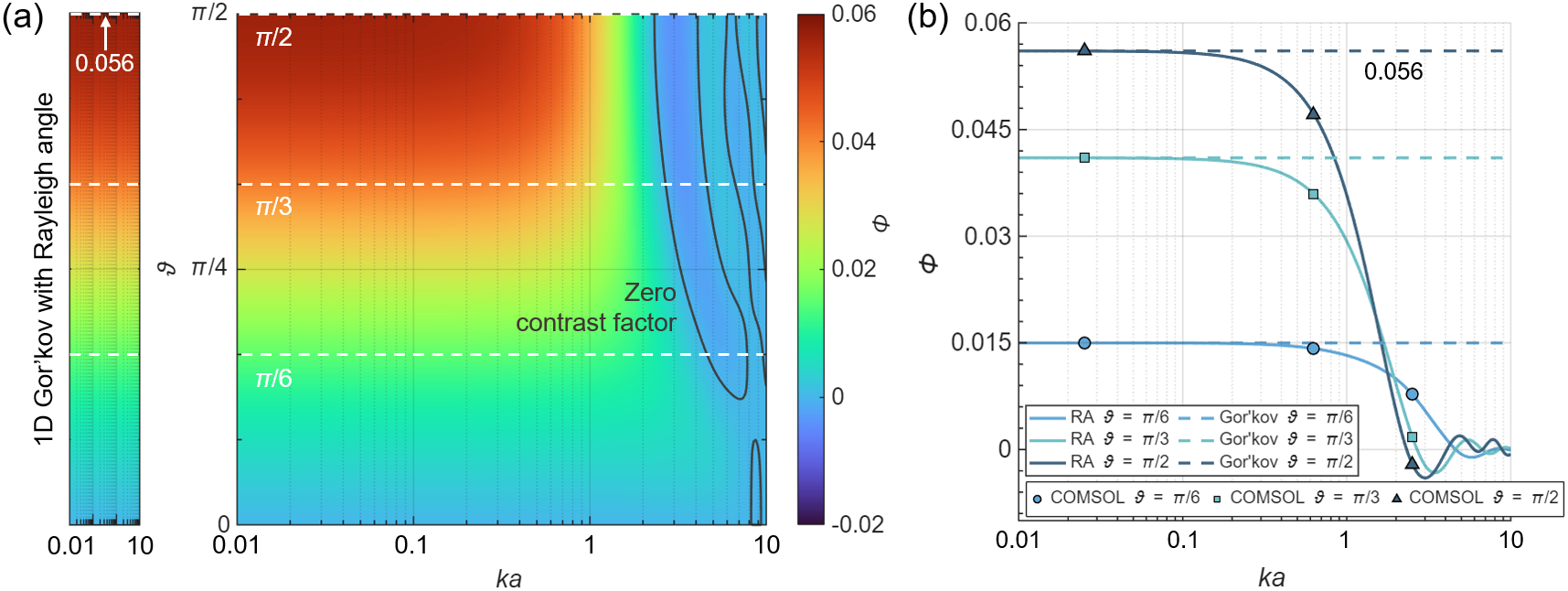}}
  \caption{The acoustophoretic contrast factor $\Phi$ of a WBC as a function of the Rayleigh angle $\vartheta$ and the dimensionless size parameter $ka$: (a) colour map of $\Phi$ showing its variation with $\vartheta$ and $ka$, where the bar on the left denotes the contrast factor predicted by the 1D Gor'kov theory with Rayleigh angle, which is independent of $ka$; (b) evolution of $\Phi$ with $ka$ for $\vartheta = \pi/6$, $\pi/3$, and $\pi/2$, together with the corresponding prediction of the 1D Gor'kov solution and the COMSOL simulation results shown as scatter points.}
\label{fig:Phi}
\end{figure}

The acoustic radiation force exerted by a 1D standing wave is of great interest in acoustofluidics, since this configuration is common in acoustic-based particle separation and trapping. Setting $\epsilon=0$ in Eq.~\eqref{eq:2D pressure} yields, 
\begin{equation}
p(x,y,z) = {p_{0z}}{e^{ik_0y\cos \vartheta }}\cos ({k_0}z\sin \vartheta ).
\label{eq:1D pressure}
\end{equation}
Then, applying Eqs.~\eqref{eq:mother equation} and~\eqref{eq:Q RA}, together with the expression for $Q_{\varphi}$ given in Appendix A, the acoustic radiation force considering scattering with higher-order harmonics along the $z$-direction for a 1D standing wave can be obtained as
\begin{equation}
\begin{aligned}
F_{acz} &= 4\pi \Phi k_0 a^3 E_{ac} \sin(2k_0z \sin\vartheta),\\
\Phi &=
-\frac{1}{(ka)^3}
\sum_{n,m}
\left[
(n-m+1)
P_n^m(\sin \vartheta)
P_{n+1}^m(\sin \vartheta)
\right] \\
&\quad \times
\cos[\pi(n-m)]
\frac{(n-m)!}{(n+m)!}
\mathrm{Im}[S_n].
\end{aligned}
\label{eq:1D}
\end{equation}
Eq.~\eqref{eq:1D} gives the analytical expression for the acoustophoretic contrast factor $\phi$ in the 1D case, while taking both high frequency and Rayleigh-angle effects into account. This factor depends on the dimensionless size parameter $ka$, incorporated into the scattering factor $S_n$ with $k = k_0$, as well as on the Rayleigh angle $\vartheta$. Owing to the complexity of its explicit expression, reference values of $\phi$ for WBC are presented directly in Fig.~\ref{fig:Phi}. 

The corresponding contrast factor using Gor'kovs theory is given by ~\citep{riaud2020mechanical},

\begin{align}
\Phi^{Gok} = \left( \frac{1}{3}f_1 - \frac{1}{2} f_2 \cos(2 \vartheta) \right)\sin\vartheta, \\
f_1 = 1- \frac{\kappa_p}{\kappa_0}, \, \kappa_0 = \frac{1}{\rho_0 c_0^2}, \,f_2= \frac{2(\rho_p-\rho_0)}{2 \rho_p + \rho_0},
\end{align}
where $f_1$ and $f_2$ denote the monopole and dipole scattering coefficients, respectively. It should also be noted that both contrast factors above are independent of the particle position in the acoustic field.

The colour map in Fig.~\ref{fig:Phi}(a) shows that the contrast factor is strongly influenced by both the parameter $ka$ and the Rayleigh angle $\vartheta$, where the black solid line denotes the contour $\Phi=0$. The variation of the contrast factor shown in the figure stands in sharp contrast to the 1D Gor'kov solution with Rayleigh angle for liquid particles~\citep{riaud2020mechanical}, in which the contrast factor is independent of $ka$, as indicated by the bar on the left. Fig.~\ref{fig:Phi}(b) presents, from panel (a), the evolution of the contrast factor with $ka$ for three representative Rayleigh angles (solid lines), $\pi/6$, $\pi/3$, and $\pi/2$, together with the corresponding solution of the 1D Gor'kov theory with Rayleigh angle (dash lines). To further enhance the credibility of the comparison, panel (b) also includes COMSOL simulation results at three representative values of $ka$, from low to high, for each Rayleigh angle. It can be seen that, when $ka$ is small, the curves calculated by the RA framework for all three Rayleigh angles remain nearly constant as $ka$ increases, indicating that the dynamical contribution of higher-order scattering modes is weak in this regime. By contrast, when $ka$ becomes large, the values of $\Phi$ for all three curves decrease rapidly and exhibit oscillations, revealing the onset of complex higher-order scattering effects. However, the predictions of the RA framework remain in good agreement with the simulation results over the entire range of $ka$ considered. Overall, the value of $\Phi$ predicted by the present RA framework is essentially identical to the Gor'kov solution when $ka$ is small; however, as $ka$ increases, the Gor'kov solution rapidly deviates from both the RA framework prediction and the simulation results. Specifically, the 1D Gor'kov solution neglecting the Rayleigh angle corresponds to $\vartheta=\pi/2$. Accordingly, it can be seen that, for the case of $\vartheta=\pi/2$, both the RA framework and the Gor'kov solution predict $\Phi \approx 0.056$ when $ka$ is small.

\subsection{2D standing waves: equilibrium points and stability}

\begin{figure}[!h]
  \centerline{\includegraphics[width=1\textwidth]{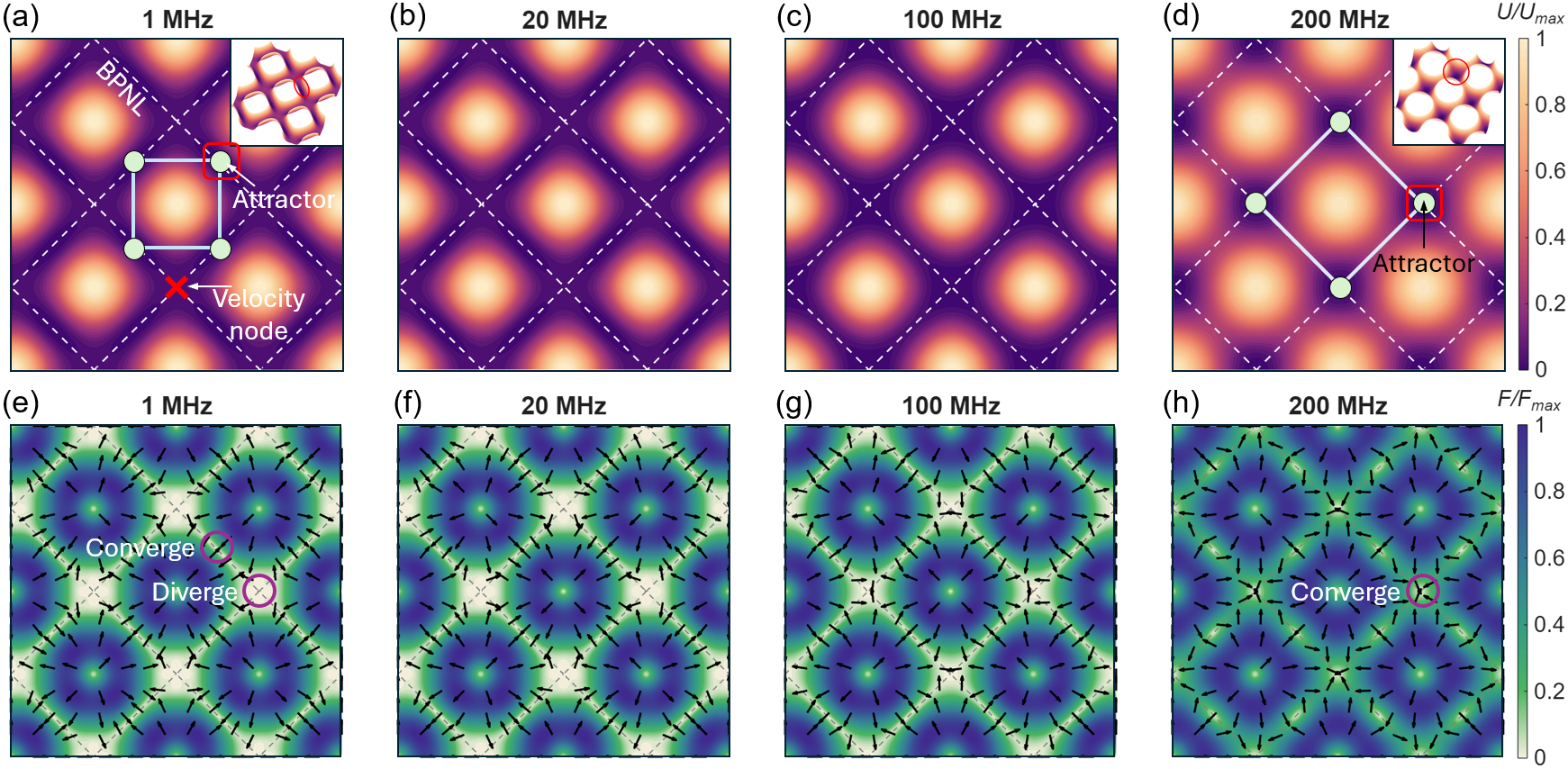}}
  \caption{Frequency-dependent evolution of the acoustic potential and acoustic radiation force field for white blood cells (WBC). (a-d) Distributions of the acoustic potential at $f = 1$, $20$, $100$, and $200$ MHz, respectively. (e-h) Corresponding distributions of the acoustic radiation force at the same frequencies. In all panels, the dashed lines denote the pressure nodal lines (BPNLs) of the 2D standing-wave field, and their intersections correspond to the velocity nodes of the background acoustic field. Panel (a) shows that, in the low-frequency regime, the minima of the acoustic potential, i.e. the attractors in the field, are located at the midpoints between adjacent velocity nodes, and the resulting attractor network exhibits a square pattern. As the frequency increases, this pattern evolves continuously. By $f = 200$ MHz, as shown in panel (d), the potential minima have shifted onto the velocity nodes themselves, and the attractor network correspondingly assumes a diamond pattern, signaling a topological transition of the acoustic potential landscape. The arrows in panels (e-h) indicate the direction of the acoustic radiation force, while the background contours represent its magnitude. Together, they show that at low frequency (1 MHz) the force diverges from the velocity nodes but converges toward the midpoints between adjacent velocity nodes, directly confirming that the particle equilibrium positions lie at these midpoints. At high frequency (200 MHz), the situation is reversed: the force converges toward the velocity nodes, indicating that the particles can be stably trapped there.}
\label{fig:Potential}
\end{figure}

From an application-oriented perspective, the goal of studying acoustofluidic systems is to control particle motion. This, in turn, requires an analysis of the equilibrium positions and their stability. Since the RA framework developed in this work enables the calculation of the acoustic radiation force in the high-frequency regime, and the acoustic radiation force is related to the potential through $\mathbf{F}=-\nabla U$, the potential field $U$ can be reconstructed from the computed radiation-force field. Using the acoustic parameters listed in Table.~\ref{tab:model_parameters}, we reconstruct the potential field for the WBC at four representative frequencies, namely 1, 20, 100, and 200 MHz. The resulting potential fields, normalized by their peak value $U_{max}$, are shown in Fig.~\ref{fig:Potential}(a–d). In these panels, the white dashed lines denote the background pressure nodal lines (BPNLs) of the 2D standing-wave field, and all of their intersections belong to the velocity nodes of the acoustic field.

Figure~\ref{fig:Potential}(a–d) reveals a frequency-dependent topological transition of the potential landscape that directly governs particle stability. According to the stability theory of dynamical systems, particles settle at locations where the potential attains local minima. At $f$ = 1 MHz, the inset in Fig.~\ref{fig:Potential}(a), which shows the corresponding three-dimensional potential landscape, indicates that the minima of the potential field lie within an elongated spindle-shaped low-potential region between two neighboring potential peaks. In other words, the attractors of the field, namely the particle equilibrium points, are located at the midpoints between adjacent velocity nodes along the BPNLs. Around each potential peak, these attractors can be connected to form a square, and we therefore refer to this topological configuration of the potential field as the ‘square mode’. As the frequency increases, the topology of the potential landscape evolves. As shown in Fig.~\ref{fig:Potential}(d), when $f$ = 200 MHz, the minima of the potential landscape shift onto the velocity nodes themselves, and the four neighboring minima around each potential peak are connected in a ‘diamond pattern’. This indicates that the topology of the potential field undergoes a transition from the square mode to the diamond mode. Such a transition is accompanied by a redistribution of the attractors in the acoustic landscape, and is therefore directly relevant to the trapping and organization of cells and microparticles in acoustofluidic systems. The phenomenon has previously been observed experimentally and numerically~\citep{collins2015two,baasch2018acoustofluidic,deng2023rectangular}, and has been shown analytically for solid spheres~\citep{silva2019particle}. The analytical confirmation for liquid particles and the inclusion of the Rayleigh angle are novelties of the underlying work.

Figures~\ref{fig:Potential}(e–h), which display the acoustic radiation force normalized by its maximum value, provide a more direct dynamical picture of how stability changes with frequency. The arrows indicate the local direction of the acoustic radiation force. In the low-frequency case, $f$ = 1 MHz, the acoustic radiation force vanishes at the velocity nodes, while the surrounding force vectors diverge outward from these points, showing that the velocity nodes are repellers. By contrast, at the midpoints between adjacent velocity nodes, the force vectors converge inward, indicating stable particle equilibrium there. As the frequency increases, this situation reverses. In the high-frequency case of $f$ = 200 MHz, the force vectors converge toward the velocity nodes, demonstrating that these nodes have become attractors of the acoustic landscape.

The frequency modifies $ka$, thereby altering the relative strengths of the different scattering orders and hence reshapes the distribution of attractors in the acoustic landscape. To understand the transition behavior of the acoustic potential field, it is essential to clarify the interplay among the particle compressibility $\kappa_p$, density $\rho_p$, and the governing parameter $ka$. Motivated by this consideration, we next introduce two probe particles to investigate this relationship in a systematic manner.

\begin{figure}[!h]
  \centerline{\includegraphics[width=0.9\textwidth]{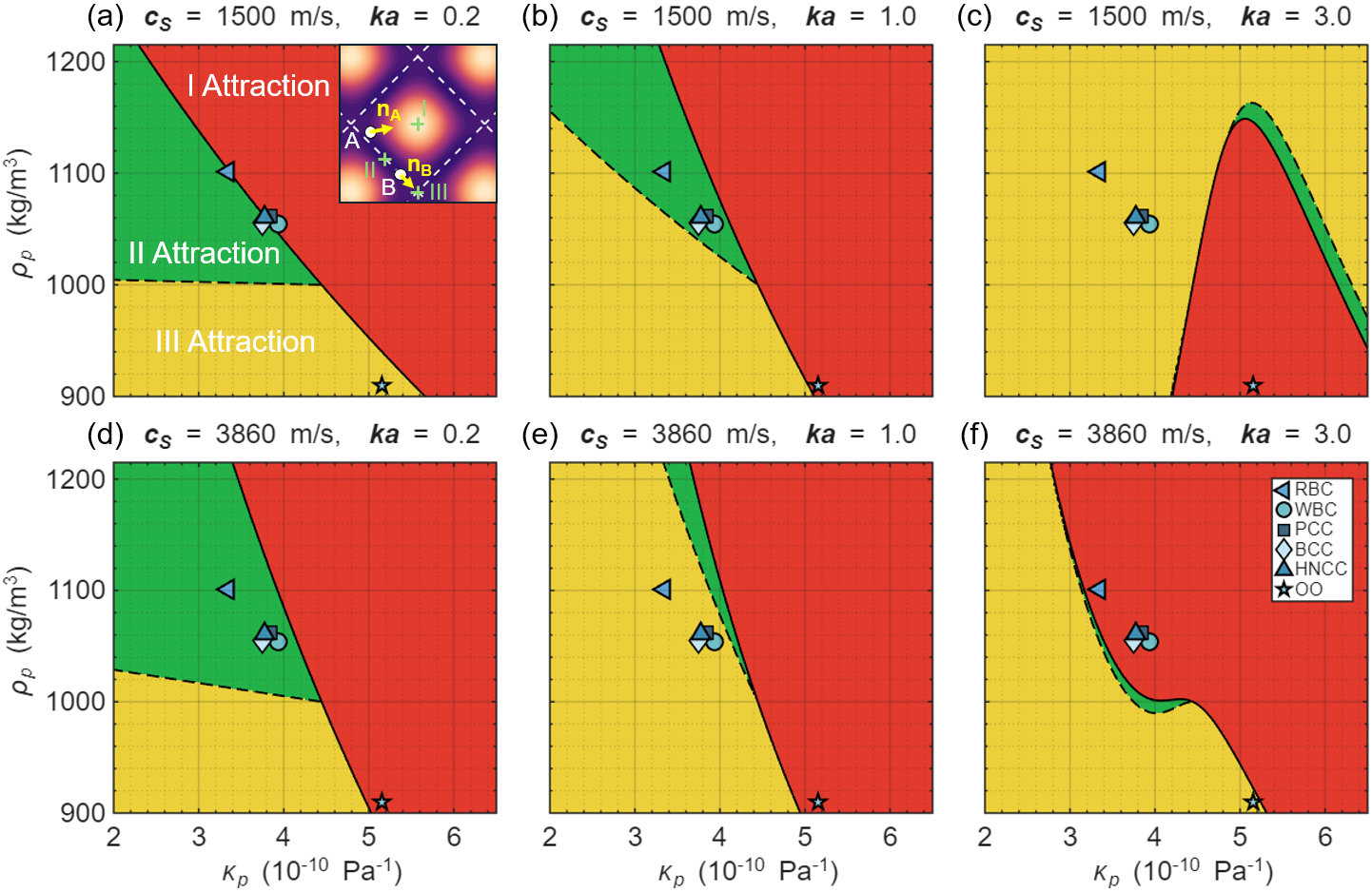}}
  \caption{Phase diagrams of particle equilibrium points in the $(\rho_p,\kappa_p)$ plane, showing the dependence of the equilibrium state on the particle density $\rho_p$ and compressibility $\kappa_p$. Panels (a-f) correspond to different combinations of the SAW device substrate sound speed $c_S$ and the parameter $ka$. Positions I, II, and III denote the three possible equilibrium points of a particle in the acoustic field. To identify which equilibrium point is realized for a given combination of $\rho_p$ and $\kappa_p$, two probe particles, A and B, are introduced at positions $(-3\lambda_S/8,\,-\lambda_S/16)$ and $(-\lambda_S/8,\,-3\lambda_S/16)$, respectively. A unit vector $\mathbf{n}_{\mathrm{A}}$ is defined at the location of particle A, pointing towards position I, while a unit vector $\mathbf{n}_{\mathrm{B}}$ is defined at the location of particle B, pointing towards position III. The acoustic radiation forces acting on particles A and B are denoted by $\mathbf{F}_{\mathrm{pA}}$ and $\mathbf{F}_{\mathrm{pB}}$, respectively. If $\mathbf{F}_{\mathrm{pA}} \cdot \mathbf{n}_{\mathrm{A}} > 0$, the equilibrium point is identified as position I; otherwise, it is either position II or position III. A further criterion is then applied: if $\mathbf{F}_{\mathrm{pB}} \cdot \mathbf{n}_{\mathrm{B}} > 0$, the equilibrium point is position III; otherwise, it is position II. The red, green, and yellow regions indicate that the particle equilibrates at positions I, II, and III, respectively. The six markers denote red blood cell (RBC), white blood cell (WBC), prostate cancer cell (PCC), breast cancer cell (BCC), head and neck squamous cancer cell (HNCC), and olive oil droplet (OO).}
\label{fig:Attractor_shift}
\end{figure}

As illustrated in the inset of Fig.~\ref{fig:Attractor_shift}(a), there are three kinds of locations in the acoustic field that may act as attractors, at which particles can remain in equilibrium. The first are the pressure antinodes, denoted as position I. The second are the midpoints between adjacent velocity nodes on the BPNLs, denoted as position II. The third are the velocity nodes on the BPNLs themselves, denoted as position III. To determine which of these three kinds of positions serves as the attractor under different physical conditions, we introduce two probe particles. Probe particle A is placed at $(-3\lambda_S/8,\,-\lambda_S/16)$, while probe particle B is placed on the BPNL at $(-\lambda_S/8,\,-3\lambda_S/16)$, with $\lambda_s=c_s/f$. The acoustic radiation forces acting on particles A and B are denoted by $\mathbf{F}_{\mathrm{pA}}$ and $\mathbf{F}_{\mathrm{pB}}$, respectively. We further define two unit vectors: $\mathbf{n}_{\mathrm{A}}$, pointing from particle A toward its closest position I, and $\mathbf{n}_{\mathrm{B}}$, pointing from particle B toward its closest position III. Under this setup, if position I is the attractor of the field, then $\mathbf{F}_{\mathrm{pA}} \cdot \mathbf{n}_{\mathrm{A}} > 0$; otherwise, the attractor must be either position II or position III. If position I is not the attractor, we then examine $\mathbf{F}_{\mathrm{pB}} \cdot \mathbf{n}_{\mathrm{B}}$ for particle B. If this quantity is positive, position III is the attractor; if it is negative, the attractor is located at position II.

Using water as the background fluid, we varied the density $\rho_p$ and compressibility $\kappa_p$ of the probe particles and constructed phase diagrams characterizing the stability of the acoustic field. In these diagrams, the red, green, and yellow regions denote parameter combinations of $\rho_p$ and $\kappa_p$ for which the attractor is located at position I, II, and III, respectively. Three values of $ka$, namely 0.2, 1.0, and 3.0, were considered, together with two substrate sound speeds of the SAW device which characterizes the Rayleigh angle, $c_S$ = [1500, 3860] m/s, corresponding respectively to the case where the Rayleigh-angle effect is neglected and the case where the substrate material is lithium niobate. In total, six phase diagrams were obtained, as shown in Fig.~\ref{fig:Attractor_shift}.

It can be seen in panel (a) that variations in $\rho_p$ and $\kappa_p$ lead to pronounced changes in the equilibrium points of the particles in the field. In general, particles with relatively large $\rho_p$ and $\kappa_p$ tend to equilibrate at position I, whereas particles with relatively small $\rho_p$ and $\kappa_p$ tend to equilibrate at position III. This switching of equilibrium behaviour within the same acoustic field originates from the change in scattering characteristics caused by variations in particle inertia and compressibility. From panel (a) to panel (b), $ka$ increases from 0.2 to 1.0; correspondingly, the red and green regions shrink, while the yellow region expands, indicating that under higher-frequency conditions particles are more likely to stabilize at position III (the velocity node), consistent with the observations in Fig.~\ref{fig:Potential}. The region pattern in panel (c) differs markedly from those in panels (a) and (b), indicating that when $ka$ becomes very large, the scattering behaviour in the acoustofluidic system has become highly complex. Panels (d)-(f) further incorporate the Rayleigh-angle effect, see panels (a)-(c). It is found that the presence of the Rayleigh-angle effect can significantly alter the equilibrium states of particles in the acoustic field. In particular, the green region decreases, indicating that, once the Rayleigh-angle effect is taken into account, particles are more likely to move towards position I and position III. A dedicated discussion of the quantitative influence of the Rayleigh-angle effect on the acoustic radiation force will be given in the later Discussion section.

In Fig.~\ref{fig:Attractor_shift}, we added six representative liquid particles commonly encountered in experiments~\citep{cushing2017ultrasound,shi2025effects}, namely red blood cell (RBC), white blood cell (WBC), DU-145 prostate cancer cell (PCC), MCF-7 breast cancer cell (BCC), LU-HNSCC-25 head and neck squamous cancer cell (HNCCs), and olive oil droplet (OO). It can be clearly seen that the equilibrium points of these particles differ substantially among the various panels. Therefore, in practical applications, this switching of particle equilibrium points with $ka$ and the Rayleigh-angle-related conditions may be exploited to achieve the separation of different particles.

\section{Discussion}

In this section, we focus on two issues of central importance in SAW-based acoustofluidic technology: first, under what range of parameters widely used Rayleigh-limit methods, such as the Gor'kov framework, remain applicable; and second, what quantitative influence the Rayleigh-angle effect intrinsic to SAW devices exerts on the acoustic radiation force.

\subsection{The application range of Rayleigh-limit approximation}

As shown in Fig.~\ref{fig:Phi}, in the 1D case the parameter $ka$ has a significant influence on the validity of the Gor'kov solution, and an upper limit $\zeta$ of $ka$ exists for which the Gor'kov solution remains applicable. We therefore turn in this section to the effect of higher-order harmonics of the scattering on the acoustic radiation force. To this end, we first present the expression for the radiation-force efficiency coefficients in a 2D standing-wave field including the Rayleigh-angle effect within the Gor'kov framework. Since Eq.~\eqref{eq:2D pressure} already gives the acoustic pressure field, the corresponding fluid velocity field can be obtained from the momentum equation of the fluid. The Gor'kov potential can then be calculated~\citep{bruus2012acoustofluidics,gor1962forces}, and, by taking its negative gradient, one obtains a more compact expression for $Q_x$ and $Q_z$ under the Gor'kov framework:

\begin{equation}
\begin{array}{l}
{Q_x} = \epsilon Q_1^{Gok}\sin(k_0x\sin\vartheta)\cos(k_0z\sin\vartheta) + \epsilon^2 Q_2^{Gok}\sin(2k_0x\sin\vartheta),\\
{Q_z} = \epsilon Q_1^{Gok}\sin(k_0z\sin\vartheta)\cos(k_0x\sin\vartheta) + Q_2^{Gok}\sin(2k_0z\sin\vartheta),
\end{array}
\label{eq:Q Gor'kov}
\end{equation}
with the corresponding coefficients are given by
\begin{equation}
\left[ \begin{array}{l}
Q_1^{Gok}\\
Q_2^{Gok}
\end{array} \right] = \frac{{2a{k_0\sin\vartheta}}}{3}\left[ \begin{array}{l}
2({f_1} - \frac{3}{2}{f_2}{\cos ^2}\vartheta )\\
{f_1} - \frac{3}{2}{f_2}\cos 2\vartheta 
\end{array} \right].
\end{equation}
In practical applications, when the acoustofluidic system remains within the Rayleigh limit, the acoustic radiation force can be calculated directly using the more compact expression given in Eq.~\eqref{eq:Q Gor'kov}.

\begin{figure}
  \centerline{\includegraphics[width=0.95\textwidth]{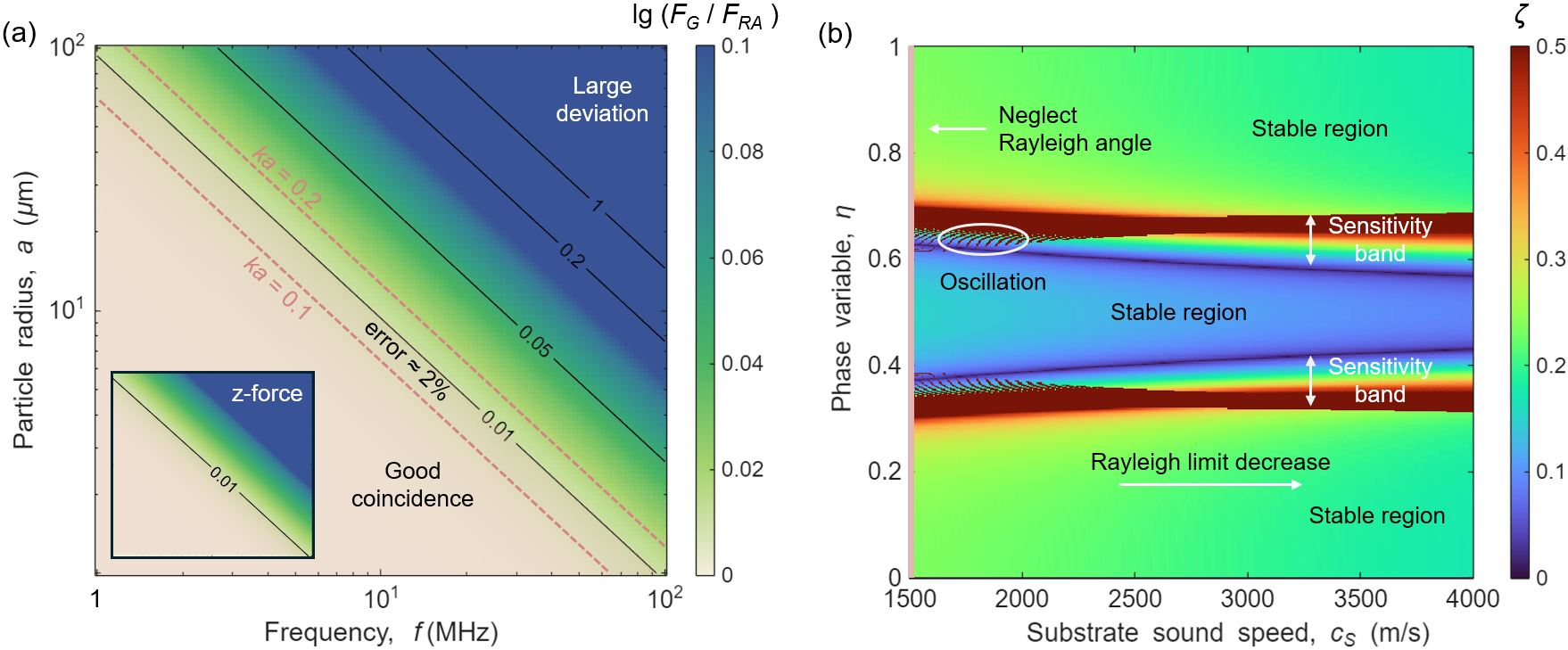}}
  \caption{Validity of Gor'kov framework: (a) Contour maps of $F_G/F_{RA}$ as functions of the acoustic frequency $f$ and particle radius $a$, with $c_S$ = 3860 m/s. Here, $F_G$ and $F_{RA}$ denote the acoustic radiation forces predicted by the Gor'kov framework and our framework, respectively. The main panels show the $x$-component of the force, while the insets show the $z$-component. As $ka$ increases, the ratio $F_G/F_{RA}$ departs progressively from unity, indicating an increasing discrepancy between the Gor'kov and RA predictions. Taking a $2\%$ deviation between $F_G$ and $F_{RA}$ as the criterion, the Rayleigh-limit threshold of the present acoustic system is estimated to be $\zeta \approx 0.1 - 0.2$. (b) Variation of the threshold $\zeta$ for the validity of Gor'kov framework with the substrate sound speed $c_S$ and the phase variable $\eta$. The threshold $\zeta$ is defined as the value of $ka$ at which the deviation between the predictions of the Gor'kov framework and the RA framework reaches $2\%$. The phase variable $\eta = xf/c_S$ is defined such that $\eta \in [0,1]$ corresponds to motion over one full wavelength. Along the $\eta$-direction, the contour map can be divided into five characteristic regions according to the intensity of the variation in $\zeta$: a stable region, a sensitivity band, a stable region, a sensitivity band, and a stable region. In particular, for small $c_S$, pronounced oscillations of the threshold can be observed inside the two sensitivity bands. These indicate that the position of the particle is itself an important factor governing the effectiveness of Gor'kov solution. As $c_S$ increases, the Rayleigh-angle effect becomes progressively more pronounced, while $\zeta$ of the acoustofluidic system decreases correspondingly, implying an enhanced contribution from higher-order scattering harmonics.}
\label{fig:Contour maps of F ratio} 
\end{figure}

In order to study the dynamical effect of higher-order harmonics scatterings, the particle is placed at the position of $(0.617\lambda_x,\,0.363\lambda_z)$. With $c_S$ = 3860 m/s, the acoustic frequency $f$ is varied from 1 to 100 MHz, while the particle radius $a$ is varied from 1 to 100 $\mathrm{\mu m}$. The acoustic radiation forces predicted by the RA framework and the Gor'kov framework, denoted by $F_{RA}$ and $F_G$, respectively, are then computed for each parameter combination. The resulting contour maps of the force ratio $F_G/F_{RA}$ in logarithmic scale are shown in Fig.~\ref{fig:Contour maps of F ratio}(a), where the main panels correspond to the $x$-component of the force and the insets to the $z$-component. It is evident that the force characteristics in the $x$- and $z$-directions are nearly identical. As in the 1D case, good agreement between the Gor'kov framework and the RA framework is observed in the 2D case when both $f$ and $a$ are small, i.e. when $ka$ is small, indicating that the acoustic radiation force can be described adequately by the monopole and dipole scattering terms alone. However, once $f$ and $a$ become sufficiently large, the discrepancy between the Gor'kov framework and the RA framework increases markedly, indicating that higher-order scattering harmonics become significant. If a relative error of $2\%$ between $F_G$ and $F_{RA}$ is adopted as the threshold for the validity of Gor'kov solution, then the upper limit of $ka$ is approximately in the range $0.1 - 0.2$.

The particle position in the acoustic field, together with the presence of the Rayleigh-angle effect, can alter the scattering characteristics of the acoustofluidic system, influence the excitation of higher-order scattering harmonics, and thereby affect the applicability limit of the Gor'kov framework. To further quantify this effect, we define the phase variable for the particle as $\eta = xf/c_S$, such that $\eta \in [0,1]$ corresponds to motion over one full wavelength. The WBC listed in Table~\ref{tab:model_parameters} is then taken to travel from the origin along the positive $x$-direction over one unit of $\eta$ in the 2D standing-wave field at $f = 10~\mathrm{MHz}$. We impose a $2\%$ deviation between the predictions of the Gor'kov framework and the RA framework as the criterion, and then solve inversely for the particle radius $a$, from which the corresponding value of $ka$ is obtained. This value of $ka$ is taken to define the threshold $\zeta$ for the validity for Gor'kov framework. The resulting contour map of $\zeta$ for different values of $c_S$ is shown in Fig.~\ref{fig:Contour maps of F ratio}(b). Overall, the threshold lies between 0.1 and 0.3 over most of the parameter space shown in the figure. Two notable features merit particular attention:

For particle positions, from an overall perspective, regardless of the value of $c_S$, the distribution of $\zeta$ with increasing phase $\eta$ exhibits a clear sequence of five characteristic regions: stable region → sensitivity band → stable region → sensitivity band → stable region. In the stable regions, $\zeta$ varies smoothly and moderately with $\eta$, whereas in the sensitivity bands it changes dramatically. This indicates that the threshold is governed not only by the physical properties of the acoustofluidic system and the particle, but also by the particle positions within the acoustic field. Near the half-wavelength positions, $\zeta$ exhibits distinctly low values. The existence of these sensitivity bands shows that, at specific positions, the scattering dynamics become extremely sensitive to variations in $ka$. A rapid increase in $\zeta$ with $\eta$ signifies that the contribution of higher-order scattering to the acoustic radiation force is sharply weakened, so that lower-order scattering dominates and $\zeta$ increases accordingly; a rapid decrease in $\zeta$ indicates the opposite trend. Notably, the locations of the sensitivity bands appear to be largely insensitive to $c_S$, as the bands are nearly horizontal in the map. For all values of $c_S$ considered, they remain concentrated around $\eta \approx 0.32-0.43$ and $\eta \approx 0.57-0.68$. In particular, when $c_S$ is relatively small, pronounced oscillations of $\zeta$ can be observed, suggesting a more complex scattering under low $c_S$ conditions. For the Rayleigh-angle effect, as $c_S$ increases, this effect becomes progressively stronger, while the threshold $\zeta$ decreases correspondingly. This demonstrates that the Rayleigh-angle effect facilitates the emergence of higher-order scattering dynamics.

\subsection{Impact of the Rayleigh-angle effect on acoustic radiation force}

The Rayleigh-angle effect not only influences the threshold $\zeta$ below which the Gor'kov framework remains valid, but can also directly modify the magnitude of the acoustic radiation force and thereby alter the dynamical state of the acoustofluidic system, because it can substantially change the effective wavelength in the system. However, in current analyses of 2D SAW dynamics, the Rayleigh angle is often neglected~\citep{collins2015two,silva2019particle,baudoin2020acoustic}. Therefore, we analyze the effect of Rayleigh angle on acoustic radiation forces based on the combination of Eqs.~\eqref{eq:mother equation} and~\eqref{eq:Q RA}.

\begin{figure}[!h]\centerline{\includegraphics[width=0.7\textwidth]{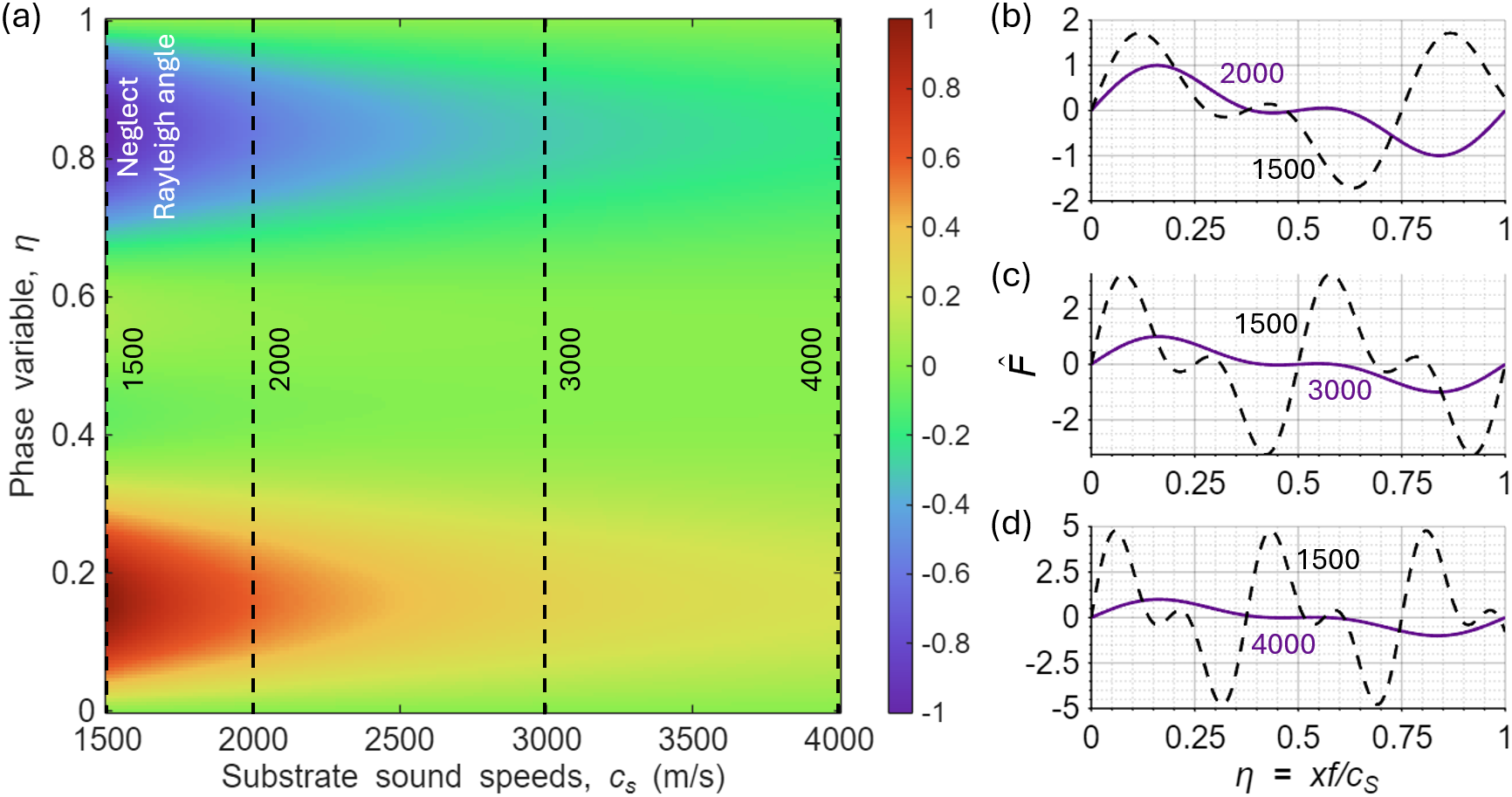}}
  \caption{Effect of the Rayleigh angle on the acoustic radiation force. (a) Phase-dependent response of the acoustic radiation force for different substrate sound speeds $c_S$ of the SAW device. (b-d) Comparison of the acoustic radiation force between $c_S = [2000, 3000, 4000]\ \mathrm{m/s}$ and $c_S = 1500\ \mathrm{m/s}$ (corresponding to the case in which the Rayleigh-angle effect is neglected), respectively. In all cases, the particle is translated from the origin (0, 0) along the positive $x$-direction. In panel (a), the acoustic radiation force is normalized by the global maximum over the entire plotted domain. In panels (b-d), the dimensionless acoustic radiation force $\hat{F}$ on the vertical axis is defined as the actual force normalized by the maximum force within one wavelength for the corresponding case with $c_S = [2000, 3000, 4000]\ \mathrm{m/s}$.}
\label{fig:EffectRayleighAngle}
\end{figure}

The particle properties of WBC listed in Table~\ref{tab:model_parameters} are selected. At a fixed frequency of $f$ = 10 MHz of 2D standing-wave, the particle is translated from the origin (0, 0) for $\eta$ ranging from 0 to 1 along the positive $x$-direction. The evolution of the acoustic radiation force $F_x$ acting on it is computed using the RA framework. The contour map of the $x$-component of the acoustic radiation force as a function of the substrate sound speed $c_S$ and phase $\eta$ is shown in Fig.~\ref{fig:EffectRayleighAngle}(a), where the force has been normalized by the global maximum over the entire domain. It can be seen that, for relatively small $c_S$, the acoustic radiation force exhibits pronounced variations as $\eta$ increases, whereas for larger $c_S$ these variations are markedly weakened, indicating a weaker and more homogeneous force distribution. This behaviour originates from the fact that, as $c_S$ increases, the Rayleigh angle becomes smaller, which reduces the $x$-component of the wavenumber within the acoustofluidic medium and correspondingly increases its wavelength. As a result, the acoustic radiation force becomes weaker and varies more smoothly.

Neglecting the Rayleigh angle in the analysis of SAW-driven dynamics is equivalent to setting $c_S$=1500 m/s, i.e. the sound speed of water. To quantify the effect of such simplification, we further compare the acoustic radiation force for $c_S$ = [2000, 3000, 4000] m/s against the case $c_S$ = 1500 m/s, as shown in Fig.~\ref{fig:EffectRayleighAngle}(b–d). For reference, lithium niobate has $c_S \approx$ 3860 m/s, which lies within the range of $c_S$ considered here. In each panel, the force curves are normalized by the maximum acoustic radiation force across a unit range of $\eta$ for the corresponding case $c_S$ = 2000, 3000, or 4000 m/s, yielding a dimensionless force $\hat{F}$. The results show that neglecting the Rayleigh angle leads to substantial discrepancies from the true response in both the phase and the magnitude of the force. These discrepancies arise not only from the modification of the wavenumber in the $x$-direction caused by the Rayleigh-angle effect, but also from the corresponding change in the effective wavelength within the acoustic field. Since any change in the substrate sound speed $c_S$ alters the effective wavelength and hence the particle phase in the wave, even if the particle remains at the same absolute position. Moreover, the larger the substrate sound speed $c_S$ of the SAW device, the greater the resulting error. When $c_S$ = 4000 m/s, which is close to the sound speed of lithium niobate, a common SAW substrate material, the error in the computed acoustic radiation force reaches nearly a factor of five. These results demonstrate that the Rayleigh-angle effect must be taken into account in any reliable analysis of SAW-induced dynamical behaviour. They also help explain the experimentally observed lower acoustophoresis performance of SAW devices relative to BAW (bulk acoustic wave) devices, in which the acoustic field is generated by chamber resonance and the Rayleigh-angle effect is absent.

\section{Conclusion}

In conclusion, by incorporating higher-order scattering harmonics of liquid particles together with the Rayleigh-angle effect, this work establishes a theory for acoustic radiation forces in 2D standing-wave fields beyond the Rayleigh limit, and further provides a general expression for the acoustophoretic contrast factor in the 1D case. On this basis, we clarify the dynamical roles of higher-order harmonics and the Rayleigh-angle effect, as well as their consequences for field topology and particle stability.

Our analysis quantitatively identifies $ka$ and the Rayleigh angle $\vartheta$ as the key parameters governing acoustic radiation forces in SAW devices, capturing, respectively, the frequency scale of the field, the particle size scale, and the acoustic properties of the device. Comparison with FEM simulations demonstrates that the present model remains accurate over a broad range when the Rayleigh-angle effect is included. Large $ka$ and large substrate sound speed $c_S$ (equivalently, small $\vartheta$) promote higher-order harmonics, leading to a pronounced reduction in the contrast factor, and neglecting the Rayleigh-angle effect can overestimate the acoustic radiation force by nearly a factor of five. For a given particle, although the validity of the Gor'kov framework only considering the low-order scatterings depends on both particle phase and substrate sound speed, its upper applicability limit typically lies in the range $ka \approx 0.1 - 0.3$. Interestingly, the particle phase within the wave can also markedly affect this limit: when the particle is located near half-wavelength positions, the limit decreases sharply. We also confirm that $ka$, particle density $\rho_p$, and compressibility $\kappa_p$ govern the topology of the acoustic potential landscape and the location of particle equilibrium points: In 2D standing-wave field, at large $ka$, equilibrium points shift to the velocity nodes on BPNLs, whereas particles with large $\rho_p$ and $\kappa_p$ are more likely to stabilize at pressure antinodes. These findings provide direct guidance for the precise trapping and manipulation of cells and other particles in acoustofluidic systems.

Looking ahead, future studies should focus on the coupled acoustic and hydrodynamic interactions in many-particle systems. Developing accurate force models and trajectory-resolving computational frameworks for such systems remains an important open problem in acoustofluidics.
\section{Acknowledgements}

S.H and T. B. acknowledge support from the Swedish Research Council (No. 2022-04041) and the SONOCRAFT project funded by the European Innovation and Research Council (GA: 101187842). D.A. and H.P. acknowledge support from the European Research Council (ERC) under the European Union's Horizon 2020 research and innovation program (grant agreement No. 853309, SONOBOTS); the Swiss National Science Foundation (SNSF) through the SNSF Project funding MINT 2022 (grant agreement No. 213058), Spark Grant (grant agreement No. 221285); and the ETH Research Grant (grant agreement No. ETH-08 20-1). H.P. acknowledges financial support from the China Scholarship Council (202306320299). 

\appendix

\section{Appendix}\label{appA}
The detailed intermediate expressions omitted from Sec.~\ref{sec:Q expression} are collected below for completeness. We substitute Eqs.~\eqref{eq:a_nm standing} and~\eqref{eq:sn} into Eq.~\eqref{eq:Qx and Qy} in order to obtain more explicit and physically transparent expressions for the Cartesian components $Q_x$ and $Q_z$ of the radiation-force efficiency vector. For the $x$-component $Q_x$ in Eq.~\eqref{eq:Qx and Qy}, the beam-shape coefficients (BSCs) take the following form
\begin{equation}
\left\{ \begin{array}{l}
{a_{nm}} = \sqrt {4\pi (2n + 1)\frac{{(n - m)!}}{{(n + m)!}}} {e^{i{k_y}y}}[\varepsilon P_n^m(0){i^{n - m}}\cos ({k_x}x + m\vartheta ) + P_n^m(\sin \vartheta )\cos ({k_z}z + \frac{\pi }{2}n - \frac{\pi }{2}m)],\\
a_{n + 1,m + 1}^* = \sqrt {4\pi (2n + 3)\frac{{(n - m)!}}{{(n + m + 2)!}}} {e^{ - i{k_y}y}}[\varepsilon P_{n + 1}^{m + 1}(0){i^{ - n + m}}\cos ({k_x}x + m\vartheta  + \vartheta ) + P_{n + 1}^{m + 1}(\sin \vartheta )\cos ({k_z}z + \frac{\pi }{2}n - \frac{\pi }{2}m)],\\
a_{n,-m}^* = \sqrt {4\pi (2n + 1)\frac{{(n + m)!}}{{(n - m)!}}} {e^{ - i{k_y}y}}[\varepsilon P_n^{ - m}(0){i^{ - n - m}}\cos ({k_x}x - m\vartheta ) + P_n^{ - m}(\sin \vartheta )\cos ({k_z}z + \frac{\pi }{2}n + \frac{\pi }{2}m)],\\
{a_{n + 1, - m - 1}} = \sqrt {4\pi (2n + 3)\frac{{(n + m + 2)!}}{{(n - m)!}}} {e^{i{k_y}y}}[\varepsilon P_{n + 1}^{ - m - 1}(0){i^{n + m + 2}}\cos ({k_x}x - m\vartheta  - \vartheta )\\
\hspace{4.3cm} + P_{n + 1}^{ - m - 1}(\sin \vartheta )\cos ({k_z}z + \frac{\pi }{2}n + \frac{\pi }{2}m + \pi )].
\end{array} \right.
\label{eq:anm_detal}
\end{equation}
Substituting these expressions into Eq.~\eqref{eq:Qx and Qy} and carrying out the algebra yields
\begin{equation}
\begin{array}{l}
Q_x = Q_1 \cos 2k_z z + Q_2 \sin 2k_z z + Q_3 \cos k_z z \cos k_x x + Q_4 \cos k_z z \sin k_x x \\
\hspace{1.6em} + Q_5 \sin k_z z \cos k_x x + Q_6 \sin k_z z \sin k_x x + Q_7 \cos 2k_x x + Q_8 \sin 2k_x x + Q_9, \\[8pt]
\left[
\begin{array}{l}
Q_1\\
Q_2\\
Q_3\\
Q_4\\
Q_5\\
Q_6\\
Q_7\\
Q_8\\
Q_9
\end{array}
\right]
=
-\frac{2}{(ka)^2}
\sum_{n,m}
\mathrm{Im}
\left[
\begin{array}{l}
\left(AI_{n_x}^m + AII_{n_x}^m\right)\sin A_{x1}\\[3pt]
\left(AI_{n_x}^m + AII_{n_x}^m\right)\cos A_{x1}\\[3pt]
\left(BI_{n_x}^m + BII_{n_x}^m\right)\cos B_{x1}\cos B_{x2}
+\left(CI_{n_x}^m + CII_{n_x}^m\right)\cos C_{x1}\cos C_{x2}\\[3pt]
-\left(BI_{n_x}^m - BII_{n_x}^m\right)\cos B_{x1}\sin B_{x2}
-\left(CI_{n_x}^m - CII_{n_x}^m\right)\sin C_{x1}\cos C_{x2}\\[3pt]
-\left(BI_{n_x}^m + BII_{n_x}^m\right)\sin B_{x1}\cos B_{x2}
-\left(CI_{n_x}^m + CII_{n_x}^m\right)\cos C_{x1}\sin C_{x2}\\[3pt]
\left(BI_{n_x}^m - BII_{n_x}^m\right)\sin B_{x1}\sin B_{x2}
+\left(CI_{n_x}^m - CII_{n_x}^m\right)\sin C_{x1}\sin C_{x2}\\[3pt]
\left(DI_{n_x}^m + DII_{n_x}^m\right)\cos D_{x1}\\[3pt]
-\left(DI_{n_x}^m - DII_{n_x}^m\right)\sin D_{x1}\\[3pt]
\left(AI_{n_x}^m + AII_{n_x}^m\right)
+\left(DI_{n_x}^m + DII_{n_x}^m\right)\cos D_{x2}
\end{array}
\right]
\frac{(n-m)!}{(n+m)!},
\end{array}
\label{eq:9Qx}
\end{equation}
where
\begin{equation}
\begin{aligned}
&\left\{
\begin{array}{l}
AI_{n_x}^m = \dfrac{1}{2} P_n^m(\sin \vartheta)\, P_{n+1}^{m+1}(\sin \vartheta)\, S_n, \\[4pt]
BI_{n_x}^m = \varepsilon\, P_n^m(\sin \vartheta)\, P_{n+1}^{m+1}(0)\, i^{-n+m}\, S_n, \\[4pt]
CI_{n_x}^m = \varepsilon\, P_n^m(0)\, P_{n+1}^{m+1}(\sin \vartheta)\, i^{\,n-m}\, S_n, \\[4pt]
DI_{n_x}^m = \dfrac{\varepsilon^2}{2} P_n^m(0)\, P_{n+1}^{m+1}(0)\, S_n, \\[4pt]
AII_{n_x}^m = \dfrac{1}{2} P_n^m(\sin \vartheta)\, P_{n+1}^{m+1}(\sin \vartheta)\, S_n^*, \\[4pt]
BII_{n_x}^m = \varepsilon\, P_n^m(\sin \vartheta)\, P_{n+1}^{m+1}(0)\, i^{\,n+m} (-1)^m S_n^*, \\[4pt]
CII_{n_x}^m = \varepsilon\, P_n^m(0)\, P_{n+1}^{m+1}(\sin \vartheta)\, (-i)^{\,n+m} (-1)^m S_n^*, \\[4pt]
DII_{n_x}^m = \dfrac{\varepsilon^2}{2} P_n^m(0)\, P_{n+1}^{m+1}(0)\, S_n^*,
\end{array}
\right.
\;,\;
\left\{
\begin{array}{l}
A_{x1} = \pi n - \pi m + \dfrac{\pi}{2}, \\[4pt]
B_{x1} = \dfrac{\pi}{2}n - \dfrac{\pi}{2}m, \\[4pt]
B_{x2} = m\vartheta + \vartheta, \\[4pt]
C_{x1} = m\vartheta, \\[4pt]
C_{x2} = \dfrac{\pi}{2}n - \dfrac{\pi}{2}m, \\[4pt]
D_{x1} = 2m\vartheta + \vartheta, \\[4pt]
D_{x2} = \vartheta,
\end{array}
\right.
\end{aligned}
\label{eq:Auxiliary polynomial for Qx}
\end{equation}
define the corresponding coefficient polynomials. Similarly, for the z-component $Q_z$ in Eq.~\eqref{eq:Qx and Qy}, the relevant BSCs are given by
\begin{equation}
\left\{ \begin{array}{l}
{a_{nm}} = \sqrt {4\pi (2n + 1)\frac{{(n - m)!}}{{(n + m)!}}} {e^{i{k_y}y}}[\varepsilon P_n^m(0){i^{n - m}}\cos ({k_x}x + m\vartheta ) + P_n^m(\sin \vartheta )\cos ({k_z}z + \frac{\pi }{2}n - \frac{\pi }{2}m)],\\
a_{n + 1,m}^* = \sqrt {4\pi (2n + 3)\frac{{(n - m + 1)!}}{{(n + m + 1)!}}} {e^{ - i{k_y}y}}[\varepsilon P_{n + 1}^m(0){i^{ - n + m - 1}}\cos ({k_x}x + m\vartheta ) + P_{n + 1}^m(\sin \vartheta )\cos ({k_z}z + \frac{\pi }{2}n - \frac{\pi }{2}m + \frac{\pi }{2})],
\end{array} \right.
\label{eq:anm for Qz}
\end{equation}
and substitution into Eq.~\eqref{eq:Qx and Qy} leads to
\begin{equation}
\begin{array}{l}
Q_z = Q_1 \cos 2k_z z + Q_2 \sin 2k_z z + Q_3 \cos k_z z \cos k_x x + Q_4 \cos k_z z \sin k_x x \\
\hspace{1.6em} + Q_5 \sin k_z z \cos k_x x + Q_6 \sin k_z z \sin k_x x + Q_7 \cos 2k_x x + Q_8 \sin 2k_x x + Q_9, \\[8pt]
\left[
\begin{array}{l}
Q_1\\
Q_2\\
Q_3\\
Q_4\\
Q_5\\
Q_6\\
Q_7\\
Q_8\\
Q_9
\end{array}
\right]
=
\frac{4}{(ka)^2}
\sum_{n,m}
\mathrm{Im}
\left[
\begin{array}{l}
A_{n_z}^m \sin A_{z1}\\[3pt]
A_{n_z}^m \cos A_{z1}\\[3pt]
B_{n_z}^m \cos B_{z1} \cos B_{z2} + C_{n_z}^m \cos C_{z1} \sin C_{z2}\\[3pt]
- B_{n_z}^m \cos B_{z1} \sin B_{z2} - C_{n_z}^m \sin C_{z1} \sin C_{z2}\\[3pt]
- B_{n_z}^m \sin B_{z1} \cos B_{z2} + C_{n_z}^m \cos C_{z1} \cos C_{z2}\\[3pt]
B_{n_z}^m \sin B_{z1} \sin B_{z2} - C_{n_z}^m \sin C_{z1} \cos C_{z2}\\[3pt]
D_{n_z}^m \cos D_{z1}\\[3pt]
- D_{n_z}^m \sin D_{z1}\\[3pt]
D_{n_z}^m
\end{array}
\right]
\frac{(n-m+1)!}{(n+m)!}.
\end{array}
\label{eq:9Qz}
\end{equation}
where
\begin{equation}
\begin{aligned}
&\left\{
\begin{array}{l}
A_{n_z}^m = -\dfrac{1}{2} P_n^m(\sin \vartheta)\, P_{n+1}^m(\sin \vartheta)\, S_n, \\[4pt]
B_{n_z}^m = \varepsilon\, P_n^m(\sin \vartheta)\, P_{n+1}^m(0)\, i^{-n+m-1}\, S_n, \\[4pt]
C_{n_z}^m = -\varepsilon\, P_n^m(0)\, P_{n+1}^m(\sin \vartheta)\, i^{\,n-m}\, S_n, \\[4pt]
D_{n_z}^m = \dfrac{\varepsilon^2}{2} P_n^m(0)\, P_{n+1}^m(0)\, i^{-1}\, S_n,
\end{array}
\right.
,\!
\left\{
\begin{array}{l}
A_{z1} = \pi n - \pi m, \\[4pt]
B_{z1} = \dfrac{\pi}{2}n - \dfrac{\pi}{2}m, \\[4pt]
B_{z2} = m\vartheta, \\[4pt]
C_{z1} = m\vartheta, \\[4pt]
C_{z2} = \dfrac{\pi}{2}n - \dfrac{\pi}{2}m, \\[4pt]
D_{z1} = 2m\vartheta .
\end{array}
\right.
\end{aligned}
\label{eq:Auxiliary polynomial for Qz}
\end{equation}

Although $\mathbf{Q}$ in Eqs.~\eqref{eq:9Qx} and~\eqref{eq:9Qz} contain as many as nine components, most of them vanish by virtue of symmetry and the periodicity of the trigonometric functions. As a result, only the four terms shown in Eq.~\eqref{eq:Q for the main text} remain, namely two in the $x$-direction and two in the $z$-direction. These four components are precisely the quantities required in Eqs.~\eqref{eq:Q RA}.
\begin{equation}
\left[
\begin{array}{l}
Q_\alpha\\
Q_\beta\\
Q_\gamma\\
Q_\varphi
\end{array}
\right]
=
\frac{4}{(ka)^2}
\sum_{n,m}
\left[
\begin{array}{l}
X_{n_1}^m \cos \dfrac{\pi(n-m)}{2}\sin(m\vartheta+\vartheta)
+ X_{n_2}^m \sin(m\vartheta)\cos \dfrac{\pi(n-m)}{2}\\[6pt]
X_{n_3}^m \sin(2m\vartheta+\vartheta)\\[6pt]
- Z_{n_2}^m \sin \dfrac{\pi(n-m)}{2}\cos(m\vartheta)
+ Z_{n_3}^m \cos(m\vartheta)\cos \dfrac{\pi(n-m)}{2}\\[6pt]
Z_{n_1}^m \cos\!\bigl[\pi(n-m)\bigr]
\end{array}
\right]
\frac{(n-m)!}{(n+m)!}\mathrm{Im}[S_n].
\label{eq:Q for the main text}
\end{equation}
In the above equation, six auxiliary polynomials are introduced for notational convenience,
\begin{equation}
\left\{
\begin{array}{l}
X_{n_1}^m = \varepsilon P_n^m(\sin \vartheta)\, P_{n+1}^{m+1}(0)\, i^{-n+m}, \\[4pt]
X_{n_2}^m = \varepsilon P_n^m(0)\, P_{n+1}^{m+1}(\sin \vartheta)\, i^{\,n-m}, \\[4pt]
X_{n_3}^m = \dfrac{\varepsilon^2}{2} P_n^m(0)\, P_{n+1}^{m+1}(0), \\[4pt]
Z_{n_1}^m = -\dfrac{1}{2}(n-m+1) P_n^m(\sin \vartheta)\, P_{n+1}^m(\sin \vartheta), \\[4pt]
Z_{n_2}^m = \varepsilon (n-m+1) P_n^m(\sin \vartheta)\, P_{n+1}^m(0)\, i^{-n+m-1}, \\[4pt]
Z_{n_3}^m = -\varepsilon (n-m+1) P_n^m(0)\, P_{n+1}^m(\sin \vartheta)\, i^{\,n-m}.
\end{array}
\right.
\label{eq:Auxiliary polynomial for Q in main text}
\end{equation}

\bibliographystyle{plain}
\bibliography{bibFile}

\end{document}